\documentclass[a4paper,10pt]{article}
\date{}
\begin{document}

\title{ Formation of a New Class of Random Fractals in Fragmentation with Mass Loss }
\author{{\small M. K. Hassan\footnote{Email: khassan@du.bangla.net}}  \\ {\small %%@
Department of Physics, Theoretical Physics Division, University of Dhaka, Dhaka 1000, %%@
Bangladesh}}

\maketitle

\begin{abstract}

\noindent
We consider the fragmentation process with mass loss and discuss self-similar properties %%@
of the arising structure both in time and space focusing on dimensional analysis. This %%@
exhibits a spectrum of mass exponents $\theta$, whose exact numerical values are given %%@
for which $x^{-\theta}$ or $t^{\theta z}$ has the dimension of particle size %%@
distribution function $c(x,t)$ where $z$ is the kinetic exponent. We also give explicit %%@
scaling solution for special case. Finally, we identify a new class of fractals ranging %%@
from random to non-random and show that the fractal dimension increases with increasing %%@
order and a transition to strictly self-similar pattern occurs when randomness is %%@
completely seized.   

\end{abstract}

\noindent
PACS number(s): 05.20.Dd,02.50.-r,05.40-y

\vspace{2mm}

\noindent
The kinetics of irreversible and sequential breakup of particles occurs in a variety of %%@
physical processes and has important applications in science and technology. These %%@
include erosion \cite{kn.has0}, grinding and crushing of solids\cite{kn.has1}, polymer %%@
degradation and fiber length reduction \cite{kn.has2}, breakup of liquid droplets %%@
\cite{kn.has3} etc. to name just a few. In recent years there has been an increasing %%@
interest in studying fragmentation allowing variations to increase the flexibility of %%@
the theory in matching the conditions of real phenomena such as extension to higher %%@
dimension \cite{kn.has4}, agglomerate erosion \cite{kn.has0}, mass loss \cite{kn.has5}, %%@
volume change \cite{kn.has6}, fragmentation-annihilation \cite{kn.has7}. The kinetic %%@
equation approach of fragmentation is linear in character which makes it analytically %%@
tractable for a large class of breakup kernels. This is contrary to the reverse process, %%@
describing the kinetics of coagulation whose mean field approach proposed by %%@
Smoluchowski is non-linear in character and solved for a limited choice of collision %%@
kernels. This may reflect the fact that breaking up of objects follows less constraints %%@
than its reverse process. Despite its apparent simplicity and the fact that the first %%@
work appeared more than a century ago, the fragmentation process is still producing %%@
nontrivial results. For example, only recently it has been observed that when particles %%@
are described by more than one dynamical quantity, such as size and shape, the system %%@
exhibits multiscaling as it obeys infinitely many conservation laws \cite{kn.has8}. %%@
Moreover, the resulting fragment distribution was shown to exhibit multifractality on a %%@
unique support when describing fragmentation and on one of infinitely many supports when %%@
describing stochastic Sierpinski gasket process \cite{kn.has4}. It has also been %%@
discovered that a shattering transition occurs as the subsequent generation of fragments %%@
has a shorter life time than the fragments of previous generation \cite{kn.has9}. %%@
McGrady and Ziff further showed in \cite{kn.has9} that the shattering regime produces a %%@
fractal dust with dimension $0<D_f<1$ due to mass being lost to the phase of zero sized %%@
particles. In one dimension, there is only one phase boundary for shattering transition %%@
which is identified by the singularity of kinetic exponent whereas in more than one %%@
dimension there are multiple phase boundaries \cite{kn.has4}. Shattering transition is %%@
also shown to be accompanied by the absence of scaling and self-averaging %%@
\cite{kn.has10}.

\vspace{3mm}

\noindent
If associating disorder with broken objects is the most natural thing to do, then %%@
searching for an order even in this disorder is the next natural thing. This forms part %%@
of our motivation of this work. In this letter, we consider the kinetics of %%@
fragmentation with continuous mass loss and look more at the geometric and scaling %%@
aspects of the process than merely trying to solve the equation. The scaling theory %%@
essentially provides solutions in the long-time and short-size limit when the particle %%@
size distribution function evolves to a simpler form as well as becomes independent of %%@
initial conditions \cite{kn.cheng}. In reality, the most experimental system evolves to %%@
the point where this behaviour is reached. Our aim is to search for an order and %%@
quantify the arising geometry of the pattern. In fact, there are many physical processes %%@
that provide an intriguing connection between geometry and physics such as Percolation, %%@
Diffusion Limited Aggregation (DLA), Self-Organised Criticalitiy (SOC), etc. These %%@
processes evolve according to a random process obeying some conservation laws and %%@
creating simple geometrical structures that traditional Euclidean geometry cannot %%@
describe. Like many statistical physics problems, exact solutions for the distribution %%@
of masses of such ramified or stringy objects and their geometry vis-a-vis measuring %%@
their fractal dimension by complete analytical means, are still a challenge even in one %%@
dimension.

\noindent
The evolution of particle size distribution function $c(x,t)$ for fragmentation with %%@
mass loss is
\begin {equation}
{{\partial c(x,t)}\over{\partial t}}  = -  c(x,t)\int_0^\infty F(y,x-y)dy %%@
+2\int_x^\infty dy c(y,t)F(x,y-x) \nonumber \\ 
 +  {{\partial}\over{\partial x}}(m(x)c(x,t))
\end {equation}
where $F(x,y)$ is the breakup kernel describing the rate at which a particle of size %%@
$(x+y)$ breaks into sizes $x$ and $y$. Fragmentation is a process whereby cuts are %%@
equivalent of seeds being sown on the fragmenting objects, thus producing two new %%@
segments. This immediately creates two more new ends belonging to the two different, %%@
newly-created fragments; in doing so, fragments start loosing their masses immediately %%@
(as if seeds were growing on either sides uniformly) until they encounter another seed %%@
or become dust-like thereby stopping loosing their masses. Therefore the model we %%@
consider can also mimic nucleation and growth of gap in one dimension which have some %%@
relevance in Kolmogorov-Avrami-Johnson-Mehl (KAJM) nucleation and growth processes  and %%@
space covering by growing rays \cite{kn.has}.

\noindent
We consider the breakup kernel to be $F(x,y)=(xy)^\beta (x+y)^{\lambda-1}$, for which %%@
the breakup rate $a(x)=\int_0^x F(y,x-y)dy=px^{2 \beta +\lambda}$, where %%@
$p={{(\Gamma(\beta+1))^2}\over{\Gamma(2 \beta+2)}}$. The first term on the right hand %%@
side of the equation $(1)$ reveals that $x^{-(2 \beta +\lambda)}$ bears the dimension of %%@
time and this put a strong constraint on the mass loss term. So, the dimensional %%@
consistency requires $m(x)=m x^{2 \beta +\lambda+1}$, with $m$ a positive real constant. %%@
This dimensional consistency has been ignored in all previous studies \cite{kn.has5} and %%@
$\gamma < 2 \beta+\lambda +1$ was identified as the recession regime and $\gamma > 2 %%@
\beta+\lambda +1$ as the fragmentation regime assuming $m(x)\sim m x^\gamma$. Since $x$ %%@
and $t$ are inextricably intertwined via the dimensional consistency, any of the two can %%@
be taken to be an independent parameter when the other one is expressible in terms of %%@
this. If $x$ is chosen to be the independent parameter then the spatial scaling {\it %%@
ansatz} is $
c(x,t) \sim x^{- \theta} \Phi(t/t_0(x))$, where $t_0(x)=x^{-(2 \beta+\lambda)}$. On the %%@
other hand, if $t$ is taken to be the independent parameter then the temporal scaling %%@
{\it ansatz} is $c(x,t) \sim t^{\theta z} \phi(x/x_0(t))$, with %%@
$x_0(t)=t^{-{{1}\over{2\beta+\lambda}}}$ and the kinetic exponent $z={{1}\over{2 %%@
\beta+\lambda}}$.
The parameters $t/t_0(x)=\xi$ and $x/x_0(t)=\eta$ are the dimensionless quantities and %%@
so are $\Phi(\xi)$ and $\phi(\eta)$. Consequently, $\theta$ takes the value for which %%@
$x^{- \theta}$ and $t^{\theta z}$ have the dimension of $c(x,t)$. Note that the spatial %%@
and temporal scaling solution are trivially connected via %%@
$\Phi(\eta^\gamma)=\eta^{\theta}\phi(\eta)$, where $\eta=xt^{{{1}\over{\gamma}}}$.
The mass exponent $\theta$ can only be found if the system follows some conservation %%@
laws. For example, for pure fragmentation ($m=0$) the mass or size of the system is a %%@
conserved quantity and gives $\theta=2$.  Defining the $n^{th}$ moment %%@
$M_n(t)=\int_0^\infty x^n c(x,t)dx$ and combining it with the rate equation $(1)$ for %%@
the present choice of $F(x,y)$ and $m(x)$ yields 
\begin{equation}
{{dM_n(t)}\over {dt}}=-[{{(\Gamma(\beta+1))^2}\over{\Gamma(2 \beta+2)}}-{{2 %%@
\Gamma(\beta+1)\Gamma(n+\beta+1)}\over{\Gamma(n+2 \beta+2)}}+mn]M_{n+2 \beta + %%@
\lambda}(t).
\end {equation}
The interesting feature of the above equation is that for $m > 0$, there are infinitely  %%@
many $n=D_f(\beta,m)$ values for which $M_{D_f(\beta,m)}(t)$s are conserved quantities. %%@
However, for $m=0$, there is only one conserved quantity $M_1(t)$, i.e. size or mass of %%@
the system, and this does not depend on $\beta$. We can find the $D_f(\beta,m)$ value by %%@
searching for the positive and real root of the equation
\begin{equation}
{{(\Gamma(\beta+1))^2}\over{\Gamma(2 \beta+2)}}-{{2 %%@
\Gamma(\beta+1)\Gamma(n+\beta+1)}\over{\Gamma(n+2 \beta+2)}}+mn=0
\end{equation}  
which is polynomial in $n$ of degree determined by the $\beta$ value. Substituting the %%@
temporal scaling {\it anstaz} into the definition of $M_n(t)$ gives $M_n(t)\sim %%@
t^{-(n-(\theta-1))z}\int_0^\infty \eta^n \phi(\eta)d \eta$ and demanding %%@
$M_{D_f}(\beta,m)$ be a conserved quantity immediately gives $\theta=(1+D_f(\beta,m))$, %%@
which clearly depends on $\beta$ and $m$ only if $m>0$. Owing to the random nature of %%@
the process and due to the presence of mass loss term, it is clear that when the process %%@
continues {\it ad infinitum}, it creates a distribution of points (dust) along a line at %%@
an extreme late stage. This distribution of points will inevitably be different from any %%@
known set such as strictly self-similar Cantor set, Julia set, Koch curve %%@
\cite{kn.has11}, stochastic or random Cantor set \cite{kn.has12}. To measure the size of %%@
the set created in the long time limit, we define a line segment %%@
$\delta={{M_n(t)}\over{M_{n-1}(t)}}\simeq t^{-{{1}\over{2 \beta+\lambda}}}$. We can %%@
count the number of such segments needed to cover the set and in the limit $\delta %%@
\longrightarrow 0$ (i.e.$ t \longrightarrow \infty$), the number $N(\delta)$ will simply  %%@
measure the set and appear to scale as $N(\delta) \sim \delta^{-D_f(\beta,m)}$. The %%@
exponent $D_f(\beta,m)$ is known as the Hausdorff-Basicovitch dimension of the set or as %%@
the fractal dimension which is simply the real positive root of the equation $(3)$.

\noindent
To get a physical picture of the role played by $m$, we set $\beta=0$ for the time being %%@
for which the equation $(3)$ becomes quadratic in $n$ and the real positive root is %%@
$D_f(m)=-{{1}\over{2}}(1+1/m)+{{1}\over{2}}\sqrt{(1+1/m)^2+4/m}$
when the second root is $D=-(D_f(m)+1+1/m)$. Therefore, the exponent $\theta$ is also %%@
function of $m$. The expression for $D_f(m)$ reveals that as $m$ value increases, the %%@
fractal dimension decreases very sharply and in the limit $m \longrightarrow \infty$,  %%@
$D_f(m)\longrightarrow 0$. This means that as $m$ increases the size of the %%@
corresponding arising set decreases sharply due to fast disappearance of its member. %%@
Whereas, as $m \longrightarrow 0$, $D_f(m) \longrightarrow 1$, that is we recover the %%@
full set (pure fragmentation) that describes a line. On the other hand had we kept $m$ %%@
fixed and let $p$ decreases the effect would have been the same as we observed for %%@
increasing $m$ with $p=1$ (i.e. $\beta =0$). Thus, it is the ratio between $m$ and $p$ %%@
that matters rather than their individual increases or decreases. To give a physical %%@
picture of what these results mean we define mass length relation for the object as $M_0 %%@
\sim \delta^{D_f(m)}$ and $M_e \sim \delta^d$ for the space where the object is being %%@
embedded, here $d$ describes the Euclidean space. The density of the property of the %%@
object $\rho$ then scales as 
\begin{equation}
\rho \sim \delta^{D_f(m)-d}.
\end{equation}
Note that for $m>0$, $D_f(m)$ is always less than one. It is thus clear that for a given %%@
class of set created by a specific rule, when $D_f(m)$ decreases it means that it is %%@
increasingly moving away from $d$ and hence  more and more members from the full set are %%@
removed. This in turn creates increasingly ramified or stringy objects since $D_f(m)=d$ %%@
describes the compact object with uniform density. So, we show that increasing $m/p$ %%@
ratio means that mass loss process gets stronger than the fragmentation process and vice %%@
versa.

\noindent
We now attempt to find the spatial scaling solution  for $\Phi(\xi)$. Note that the %%@
dimension of the arising pattern is independent of $\lambda$ and consequently %%@
independent of how fast or slow the system performs the process. So, we can set %%@
$\lambda=1$ without fear of missing any physics but it certainly simplifies our %%@
calculation. Substituting the spatial scaling {\it ansatz} into the rate equation $(1)$ %%@
for $F(x,y)=1$ and $m(x)=mx^2$ and differentiating it with respect to $\xi$, transforms %%@
the partial integro-differential equation into an ordinary differential equation , 
\begin{equation}
\xi(1-m \xi)\Phi^{\prime \prime}(\xi) %%@
+[(1-\theta)-\xi(2m(2-\theta)-1]\Phi^{\prime}(\xi)-(m(2-\theta)(1-\theta)
-(3-\theta))\Phi(\xi)=0.
\end{equation}
For $m=1$ this is hypergeometric differential equation \cite{kn.hass} whose only %%@
physically acceptable linearly independent solutions are $~_2F_1(1,-(1+2D_f);-D_f;\xi)$ %%@
and $\xi^{(1+D_f)}~_2F_1(2+D_f,-D_f;2+D_f;\xi)$, where $D_f=0.414213$. From these exact %%@
solutions for spatial scaling function we can obtain the asymptotic temporal scaling %%@
function $\phi(\xi) \sim e^{-D_f \xi}$ that satisfies the condition %%@
$\phi(\xi)\longrightarrow 0$ as $\xi \longrightarrow \infty$.

\noindent
We now attempt to see the role of $\beta$ on the system. To judge its role, it is clear %%@
from the previous discussion that we ought to give equal weight to all the terms in the %%@
equation $(1)$ so that each of them can compete on an equal footing. This can be done if %%@
only we set $m=p={{(\Gamma(\beta+1))^2}\over{\Gamma(2\beta+2)}}$ so that the relative %%@
strength  between fragmentation and mass loss process stays the same as $\beta$ value %%@
increases. This is a very crucial point to be emphasized. We can obtain the fractal %%@
dimension for different values of $\beta$, which is simply the real positive root of the %%@
equation $(3)$. A detailed survey reveals that the fractal dimension increases %%@
monotonically with increasing $\beta$. To find the fractal dimension in the limit $\beta %%@
\longrightarrow \infty$, we can use the Stirling's approximation in $(3)$ to obtain %%@
$\ln[n+1]=(1-n)\ln[2]$ when $n=0.4569997$ solves this equation. In order to give a %%@
physical picture of the role of $\beta$ in the limit $\beta \longrightarrow \infty$, we %%@
consider the following model $F(x,y)=(x+y)^\gamma \delta(x-y)$. This model describes %%@
that cuts are only allowed to be in the middle in order to produce two fragments of %%@
equal size at each time event. This makes $a(x)={{1}\over{2}}x^\gamma$, so we need to %%@
choose $m(x)={{1}\over{2}}x^{\gamma+1}$, where $m={{1}\over{2}}$ gives the same weight %%@
as for the fragmentation process. Then the rate equation for $M_n(t)$ becomes
\begin{equation}
{{dM_n(t)}\over {dt}}=-[{{(n+1)}\over{2}}-2^{-n}]M_{n+\gamma}(t).
\end{equation}
As before we set the numerical factor of the right hand side of this equation equal to %%@
zero and then take natural log on both sides to obtain the $n$ value for which $M_n(t)$ %%@
is time independent. In doing so, we arrive at the same functional equation for $n$ as %%@
we found for $\beta \longrightarrow \infty$. This shows that the kernel %%@
$F(x,y)=(xy)^\beta (x+y)^{\lambda-1}$ behaves exactly in the same fashion as for %%@
$F(x,y)=(x+y)^\gamma \delta(x-y)$. We thus find that in the limit $\beta \longrightarrow %%@
\infty$, the resulting distribution of points is a set with fractal dimension %%@
$D_f=0.4569997$ which is a strictly self-similar fractal as randomness is seized by %%@
dividing fragments into equal pieces. We are now in a position to give a physical %%@
picture of the role played by $\beta$. First of all, the process with $\beta=0$ %%@
describes the frequency curve of placing cuts about the size of the fragmenting %%@
particles is Poisson in nature. Consequently, the system enjoys the maximum randomness %%@
and the corresponding fractal dimension is $D_f=0.414213$. Whereas, for $\beta >0$, the %%@
frequency curve of placing cuts about the size of the fragmenting particles is Gaussian %%@
in nature meaning as $\beta$ value increases particles are increasingly more likely to %%@
break in the middle than on either end. That is, as $\beta$ increases, the variance %%@
decreases in such a manner that in the limit $\beta \longrightarrow \infty$ the variance %%@
of the frequency curve becomes infinitely narrow meaning a delta function distribution %%@
for which fragments are broken into two equal pieces. This analysis also specify that %%@
the rules determining the location where to place the cut are determined by the details %%@
of breakup kernel $F(x,y)$ rather than the breakup rate $a(x)$. So, there is a spectrum %%@
of fractal dimensions between $\beta \rightarrow 0$ when $D_f=0.414213$ and $\beta %%@
\longrightarrow \infty$ when $D_f=0.4569997$. A detailed numerical survey that we do not %%@
present here confirms that fractal dimension increases monotonically with $\beta$ and %%@
reaches to a constant value when $\beta \longrightarrow \infty$ in a similar fashion as %%@
the variation of $q$ with $t$ during charging process in $RC$ circuit. According to %%@
equation $(4)$ increasing $\beta$ vis-a-vis increasing order also means that the system %%@
losses less and less mass from the system and this happens despite the fact that now %%@
${{m}\over{p}}$ ratio stays the same.  Perhaps it is note worthy to mention that the %%@
present model with $\beta=\lambda=0$ and $m=1$ correspond to Yule-Furry processes for %%@
cosmic shower theory with collision loss \cite{kn.cos}, though there too dimensional %%@
consistency was ignored.

\vspace{2mm}

\noindent
In summary, we have identified a new set with a wide range of subsets produced by tuning %%@
the degree of randomness only. The process starts with an initiator of unit interval %%@
$[0.1]$ and the generator divide the interval into two pieces and deleting some parts %%@
from either sides of both the pieces at each time step. The amount of the parts to be %%@
deleted is determined by the parameter that control the intensity of randomness. When %%@
this operation continues {\it ad infinitum}, what remains is an infinite number of dust %%@
scattered over the interval. We quantified the size of the arising set by fractal %%@
dimension and  showed that the fractal dimension increases with increasing order and %%@
reaches its maximum value when the pattern described by the set is perfectly ordered, %%@
which is contrary to some recently found results \cite{kn.has13}. We have also shown %%@
that the increase of fractal dimension and the increase of mass exponent $\theta$ go %%@
hand in hand since they are intimately connected. To the best of our knowledge the exact %%@
numerical value of this mass exponent has never been reported. We have given a scaling %%@
description of the process both in time and space and obtained explicit scaling function %%@
for special case of interest. Finally we argue on the basis of our findings that fractal %%@
dimension, degree of order and the extent of ramifications of the arising pattern are %%@
interconnected.

\vspace{2mm}

\noindent
The author is grateful to R. M. Ziff for sending valuable comments. The author also %%@
acknowledges inspiring correspondence with P. L. Krapivsky and support from the Ministry %%@
of Science and Technology of Bangladesh under Grant No. 1/98/112/1(7).

\end{document}